\documentstyle[preprint,aps]{revtex}
\tighten
\begin{document}
\title{Critical visibility for 
N-particle Greenberger-Horne-Zeilinger 
correlations to violate local realism
}
\author{
Marek \.Zukowski and Dagomir Kaszlikowski}
\address{Instytut Fizyki Teoretycznej i Astrofizyki\\
Uniwersytet Gda\'nski, PL-80-952 Gda\'nsk, Poland}
\date{\today}
\maketitle
\begin{abstract}
A sequence of Bell inequalities for $N$-particle systems, which involve three
settings of each of the local measuring apparatuses, is derived.  For GHZ
states, quantum mechanics violates these inequalities by factors
exponentially growing with $N$.  The threshold visibilities of the
multiparticle sinusoidal interference fringes, for which local realistic
theories are ruled out, decrease as $(2/3)^{N}$.

\end{abstract}

\pacs{PACS numbers: 3.65.Bz, 42.50.Dv, 89.70.+c}


Greenberger-Horne-Zeilinger (GHZ) correlations \cite {GHZ} lead to a
strikingly more direct refutation of the argument of Einstein Podolsky and
Rosen (EPR), on the possibility of introducing elements of reality to
complete quantum mechanics \cite {EPR}, than considerations involving only
pairs of particles.  The EPR ideas are based on the observation that for some
systems quantum mechanics predicts perfect correlations of their properties.
However, in the case of three or more particles, in the entangled GHZ state,
such correlations cannot be consistently used to infer at a distance hidden
properties of the particles. In contradistinction to the original two
particle Bell theorem, the idea of EPR, to turn the exact predictions of
quantum mechanics against the claim of its completeness, breaks down already
at the stage of defining the elements of reality.

The reasoning of GHZ involved perfectly correlated particle systems.
However, the actual data collected in a real laboratory would reveal less
than perfect correlations, and the imperfections of the particle collection
systems would leave many of the potential events undetected.  Therefore the
original GHZ reasoning cannot be ever tested in the laboratory, and one is
forced to make some modifications (already, e.g., in
\cite{GHSZ}).

To face these difficulties several $N$-particle Bell inequalities appeared in
the literature \cite{MERMIN},\cite{ZUK}.
All these works show that quantum predictions for GHZ states violate these
inequalities by an amount that grows exponentially with $N$.  The increasing
number of particles, in this case, does not bring us closer to the classical
realm, but rather, makes the discrepancies between the quantum and the
classical more profound.

Any experimental realization of GHZ processes will be much more involved than
a two-particle Bell test (for specific proposals see \cite{OTHER} and
\cite{YURKE}). Therefore, the study of three or more particle 
interference effects does not seem to be a good route towards a loophole free
test of the hypothesis of local hidden variables. However, other questions
are to be answered in such experiments.  For example, is it at all possible
to observe three or four particle interferometric fringes which have no
classical model?  Before attempting such an experiment one must know the
borderline between the quantum and the classical (i.e., local realism).
According to current literature (with the exception of \cite{ZUK}) we enter
the non-classical territory when the fringes in a $N$-particle interference
experiment have visibilities higher than ${2}^{\frac{1}{2}(1-N)}$. The
principal aim of this work is to show that, if one allows each of the local
observers to have {\it three} measurements to choose form (instead of the
usual {\it two}), the actual threshold is lower (for $N>3$).

In a GHZ-Bell type experiment one has a source emitting $N$-particles each of
which propagates towards one of $N$ spatially separated measuring devices.
The generic form of a GHZ $N$-particle state is 
\begin{equation}
|\Psi(N)\rangle=\frac{1}{{\sqrt{2}}}(|+\rangle_{1}\dots|+\rangle_{N}+|-\rangle_{
1}\dots
|-\rangle_{N}).
\label{kukurydza1}
\end{equation}
Let as assume that the operation of each of the measuring apparata is
controlled by a knob which sets a parameter $\phi_{l}$, and the $l$-th
apparatus measures a dichotomic observable $O_{l}(\phi_{l})$ with two
eigenvalues $\pm 1$ and the eigenstates defined by
$|\pm,\phi_{l}\rangle_{l}=\frac{1}{\sqrt{2}}\left(|+\rangle_{l}\pm
e^{(i\phi_{l})}|-\rangle_{l}\right).$ The quantum prediction for obtaining
specific results at the $N$ measurement stations (for the idealized, perfect,
experiment) reads
\begin{eqnarray}
&P_{QM}^{(N)}(r_{1},r_{2},\dots,r_{N}|\phi_{1},\dots,\phi_{N})
&\nonumber\\
&={1\over 2^N}\left[1+\prod_{l=1}^{N}r_{l}
\cos\left(\sum_{k=1}^{N}\phi_{k}\right)\right]&,
\label{kukurydza4}
\end{eqnarray}
($r_{l}$ equal to $-1$ or $+1$).
The GHZ correlation function is defined as
\begin{eqnarray}
&E^{(N)}(\phi_{1},\dots,\phi_{N})&\nonumber\\
&=\sum_{r_{1},r_{2},\dots,r_{N}=1}\prod_{l=1}^{N}r_{l}P^{(N)}(r_{1},\dots,
r_{N}|\phi_{1},\dots,\phi_{N}),&
\label{kukurydza5}
\end{eqnarray}
and in the case of quantum mechanics, i.e. for $P^{(N)}=P_{QM}^{(N)}$, it
reads $ E_{QM}^{(N)}=\cos(\sum_{l=1}^{N}\phi_{l}).  $
\par
From the perspective of local realism one can try to 
give a more complete specification of the state of a member of the ensemble
of $N$-particle systems than the one given by $|\Psi(N)\rangle$.
The usual approach is to define a space of hidden states $\Lambda$ and
a probability distribution $\rho(\lambda)$ of such states and to represent
the probability of specific results by
\begin{eqnarray}
&P_{HV}^{(N)}(r_{1},\dots,r_{N}|\phi_{1},\dots,\phi_{N})&\nonumber\\
&=\int_{\Lambda}d\lambda\rho(\lambda)\prod_{l=1}^{N}P_{l}(r_{l}|\lambda,&
\phi_{l})
\label{kukurydza7}
\end{eqnarray}
where $P_{l}(r_{l}|\lambda,\phi_{l})$ is the probability to obtain result
$r_{l}$ in the $l$-th apparatus under the condition that the hidden state is
$\lambda$ and the macroscopic variable defining the locally
measured observable is set to the value $\phi_{l}$ \cite{FOOT1}. The locality
of this description is guaranteed by independence of $P_{l}$ on $\phi_{i}$
for all $i\neq l$.

We shall now derive a series of inequalities for the $N$-particle GHZ
processes based on the following simple geometric observation
\cite{ZUK}. 
Assume that one knows the components of a certain (Euclidean) vector $\vec{q}$
(the {\it known} vector), whereas about a second vector $\vec{h}$ (the {\it
test} vector) one is only able to establish that its scalar product with
$\vec{q}$ satisfies the inequality $\vec{h}\cdot \vec{q} < ||\vec{q}||^2=
\vec{q} \cdot \vec{q}$. 
Then $\vec{h}\neq\vec{q}$. This simple geometric theorem can be extended to
any pair of (real) objects for which one can define the scalar product (e.g.,
{\it matrices} or functions \cite{ZUK}).

We assume that each of the $N$ spatially separated observers has {three}
measurements to choose from.  The local phases that they are allowed to set
are $\phi_{1}^{1}=\pi/6,
\phi_{2}^{1}=\pi/2,\phi_{3}^{1}=5\pi/6$ 
(for the first observer) and for all the other $N-1$ observers they are
$\phi_{1}^{i}=0, \phi_{2}^{i}=\pi/3,\phi^{i}_{3}=2\pi/3$,
($i=2,\dots,N$).

Out of the quantum predictions for the $N$-particle correlation function at
these settings one can construct a matrix endowed with $N$ indices
\begin{eqnarray}
&E_{QM}^{(N)}(\phi_{i_{1}}^{1},\dots\phi_{i_{N}}^{N})=\cos\left(\sum_{k=1}^{N}
\phi_{i_{k}}^{k}\right)
=Q_{i_{1},\dots,i_{N}}^{(N)}
\label{kukurydza8}
\end{eqnarray}
($i_{k}=1,2,3$).

All that we know about local hidden variable theories is that their
predictions (for the same set of settings as above) must have the following
form:
\begin{eqnarray}
&E_{HV}^{(N)}(\phi_{i_{1}}^{1},\dots\phi_{i_{N}}^{N})& \nonumber\\
&=\int
d\lambda\rho(\lambda)\prod_{k=1}^{N}I_{k}(\lambda
,\phi_{i_{k}}^{k})=H_{i_{1},\dots,i_{N}}^{(N)},&
\label{kukurydza10}
\end{eqnarray}
where
\begin{eqnarray}
&I_{k}(\lambda,\phi_{i_{k}}^{k})=
\sum_{r_{k}}r_{k}P_{k}(r_{k}|\lambda,\phi_{i_{k}}^{k}).&
\label{kukurydza11}
\end{eqnarray}
Of course, in the case of a {\it deterministic} theory
$I_{k}(\lambda,\phi)=\pm1$ \cite{FOOT1}.  $H^{(N)}$ is our test matrix.
Please note, that all one knows about $H^{(N)}$ is its structure.

The scalar product of two real matrices is defined by
\begin{equation}
(H^{(N)},Q^{(N)})=\sum_{i_{1},\dots,i_{N}}H_{i_{1},\dots,i_{N}}^{(N)}
Q_{i_{1},\dots,i_{N}}^{(N)}.
\label{kukurydza12}
\end{equation} 
Our aim is to show the incompatibility of the local hidden variable
description with the quantum prediction.  To this end, we shall show that,
for two or more particles,
\begin{eqnarray} 
&({Q}^{(N)},{H}^{(N)})&\nonumber\\ &\leq
2^{N-1}\sqrt{3}<||{Q}^{(N)}||^2=\frac{3^N}{2}&.
\label{rw1} 
\end{eqnarray} 

First, we show that $||{Q}^{(N)}||^2=3^N/2$.  This can be reached in
the following way:
\begin{eqnarray}
&||{Q}^{(N)}||^2=\sum_{i_{1},\dots,i_{N}}
\cos^2\left(\sum_{k=1}^{N}\phi_{i_{k}}^{k}\right)&\nonumber\\
&=\frac{1}{2}\sum_{i_{1},\dots,i_{N}}
\left[1+\cos\left(2i\sum_{k=1}^{N}\phi_{i_{k}}^{k}\right)\right]
&\nonumber\\ &=Re\left\{\sum_{i_{1},\dots,i_{N}}
\left[1+\exp\left(2\sum_{k=1}^{N}\phi_{i_{k}}^{k}\right)\right]\right\}&
\nonumber\\
&=3^N/2+Re\left(\prod^{N}_{k=1}
\sum_{i_{k}=1}^{3}\exp(2i\phi_{i_{k}}^{k})\right),&
\end{eqnarray}
where $Re$ denotes the real part.  Since
$\sum_{l=1}^{3}e^{i(l-1)(2/3)\pi}=0$, the last term vanishes.

The scalar product $(H^{(N)},Q^{(N)})$ is bounded from above by the maximal
possible value of
\begin{equation}
S^{(N)}_{\lambda}=\sum_{i_{1},\dots,i_{N}}\left[
\cos\left(\sum^{N}_{k=1}\phi_{i_{k}}^{k}\right)
\prod_{l=1}^{N}I_{l}(\lambda, \phi_{i_{l}}^{l})\right],
\label{kukurydza13}
\end{equation}
and for $N\geq2$
\begin{equation}
S^{(N)}_{\lambda}\leq {2^{N-1}\sqrt{3}}.
\label{S}
\end{equation}

To show (\ref{S}), let us first notice that 
\begin{eqnarray} 
S_{\lambda}^{(N)}=Re
\left[\prod_{k=1}^{N}\sum_{i_{k}=1}^{3}I_{k}(\lambda|\phi_{i_{
k}}^{k})\exp(i\phi_{i_{k}}^{k}))\right].
\label{rw8} 
\end{eqnarray}
For $k=2,\dots,N$, one has $e^{i\phi_{l}^{k}}=e^{i[(l-1)/3]\pi}$ whereas
for $k=1$, $e^{i\phi_{l}^{1}}=e^{i(\pi/6)}e^{i[(l-1)/3]\pi}$.
Thus, since $I(\lambda|\cdot)=\pm 1$, the possible values for
\begin{equation}
z_{1}^{\lambda}=\sum_{i_{1}=1}^{3}I_{1}(\lambda|\phi_{i_{1}}^{1})
\exp(i\phi_{i_{1}}^{1})
\label{rw11}
\end{equation}
are $0$, $\pm 2e^{i\pi/2}$, $\pm 2e^{-i(\pi/6)} $, or
finally $\pm 2e^{i(\pi/6)} $, whereas for $k=2,\dots,N$ the possible
values of
\begin{equation} 
z_{k}^{\lambda}=\sum_{i=1}^{3}I_{k}(\lambda|\phi_{i_{k}}^{k})
\exp(i\phi_{i_{k}}^{k})
\label{rw12} 
\end{equation} 
have their complex phases shifted by $\pi/6$ with respect to the
previous set; i.e., they are $0$, $\pm 2e^{i(2\pi/3)}$, 
$\pm 2$, or finally $\pm 2e^{i(\pi/3)}$.  Since
$|z_{1}^{\lambda}\prod_{k=2}^{N}z_{k}^{\lambda} |\leq 2^{N}$ and the minimal
possible overall complex phase (modulo $2\pi$) of 
$z_{1}^{\lambda}\prod_{k=2}^{N}z_{k}^{\lambda}$ is 
$\pi/6$, one has $Re(z_{1}^{\lambda}
\prod_{k=2}^{N}z_{k}^{\lambda})\leq 2^{N}\cos(\pi/6).$
Thus inequalities (\ref{S}) and (\ref{rw1}) hold.

The left inequality of (\ref{rw1}) is a Bell inequality for the $N$-particle
experiment. If one replaces ${H}^{(N)}$ by the quantum prediction ${Q}^{(N)}$
(compare (\ref{kukurydza8}))
the inequality is violated since
\begin{equation}
({Q^{(N)}},{Q^{(N)}})=\frac{3^{N}}{2}>2^{N-1}\sqrt{3},
\label{row1}
\end{equation}
i.e., (\ref{rw1}) is violated by the factor
$(3/2)^N/\sqrt3$ (compare \cite{MERMIN}).

The magnitude of violation of a Bell inequality is not a parameter which is
directly observable in the experiment. It is rather the visibility of the
$N$-particle interference fringes which can be directly observed.  Further,
the significance of all Bell-type experiments depends on the efficiency of
the collection of the particles. Below a certain threshold value for this
parameter experiments cannot be considered as tests of local realism. They
may confirm the quantum predictions but are not falsifications of the
hypothesis of local hidden variables. Therefore we will search for the
critical visibility of $N$-particle fringes and collection efficiency, which
do not allow anymore a local realistic model.

In a real experiment (under the assumption that quantum mechanics gives
idealized, but correct predictions), the visibility of the $N$-particle
fringes, $V(N)$, would certainly be less than 1. Also the probability of
registering all potential events would be reduced by the overall collection
efficiency. If one assumes that all $N$ local apparata have the same
collection efficiency $\eta$, and takes into account that these operate
independently of each other, one can model the expected experimental results
by
\begin{eqnarray}
&P_{expt}^{(N)}(r_{1},\dots,r_{N}|\phi^{1},\dots,\phi^{N})&\nonumber\\
&=\eta^{N}{\left(\frac{1}{2}\right)^{N}}\left(1+V(N)\prod_{l=1}^{N}r_{l}
\cos\sum_{k=1}^{N}\phi
^{k}\right)&.
\label{24}
\end{eqnarray}

The full set of events at a given measuring station consists now of the
results $+1$ and $-1$, when we succeed to measure the dichotomic observable,
and a non-detection event (which is, in principle observable, if one uses
event-ready state preparation \cite{YURKE}) for which one can introduce the
value $0$. The local realistic description requires that the probabilities of
the possible events should be given by
\begin{eqnarray}
&P_{expt}^{HV}(m_{1},\dots,m_{N}|\phi^{1},\dots,\phi^{N})&\nonumber\\ &=\int
d\lambda\rho(\lambda)\prod_{k=1}^{N}P_{k}(m_{k}|\lambda,
\phi^{k})&,
\label{row4}
\end{eqnarray}
with $m_{i}=+1, -1$ or $0$. The local hidden
variable correlation function for the experimental results (at the chosen
settings) is now given by
\begin{equation}
{E_{expt}^{HV}}_{i_{1},\dots,i_{N}}=\int
d\lambda\rho(\lambda)\prod_{k=1}^{N}I_{k}'(
\lambda, \phi_{i_{k}}^{k}),
\label{row8}
\end{equation}
with
\begin{equation}
I_{k}'(\lambda,\phi_{i_{k}}^{k})=\sum_{m_{k}=-1,0,+1}m_{k}
P_{k}^{HV}(m_{k}|\lambda
,\phi_{i_{k}}^{k}).
\label{row9}
\end{equation}
For deterministic models one has now $I_{k}'(\lambda,\phi_{i_{k}}^{k})
=1,0,-1$.

One can impose several symmetries on ${P_{expt}^{HV}}$. These symmetries are
satisfied by the quantum prediction (\ref{24}), and we can expect them to
be satisfied in real experiments, within experimental error. The one that we
impose here is that:

For all sets of results, $\{m_{1},\dots,m_{N}\}$, that have equal number of
zeros (one zero or more) the probability $P_{expt}^{HV}(m_{1},\dots,m_{N})$ 
has
the same value, and this value is independent of the settings of the local
parameters $\{\phi_{i_{1}}^{1},\dots,\phi_{i_{N}}^{N}\}$.

One can define a function $f_{N}(m)$ which for $m=+1,-1,0$ has the following
values: $f(\pm 1)=\pm 1$, $f(0)=-1$ (compare \cite{GARG}) and introduce
auxiliary correlation function
\begin{eqnarray}
&\tilde{E}_{i_{1},\dots,i_{N}} =\int
d\lambda\rho(\lambda)\sum_{m_{1},\dots,m_{N}=-1,0,+1}&\nonumber\\
&\times\prod_{k=1}^{N}[f(m_{k})P_{k}(m_{k}|\lambda,
\phi_{i_{k}}^{k})]={\tilde{H}}^{(N)}_{i_{1},\dots,i_{N}}&.
\label{row3}
\end{eqnarray}
Since, due to the symmetry conditions, one has, e.g., $\sum_{m_{2}=1,-1}
f(m_{2})P_{expt}^{HV}(0,m_{2},m_{3},\dots,m_{N})=0$, the
following relation results:
\begin{equation}
\tilde{E}_{i_{1},\dots,i_{N}}={E^{HV}_{expt}}_{i_{1},\dots,i_{N}}+[f(0)]^
{N} P(0,\dots,0),
\label{row6}
\end{equation}
where
$P(0,\dots,0)$
is the probability that all detectors would fail to register particles,
and under our assumptions
it is independent of the settings, and equals $(1-\eta)^N$.

The auxiliary correlation function must satisfy the original inequality
(\ref{rw1}); i.e., one has
\begin{equation}
(Q^{(N)},{\tilde{H}}^{(N)})\leq{2^{N-1}}\sqrt{3}.
\label{row10}
\end{equation}
However, this implies that
\begin{eqnarray}
&-{2^{N-1}}\sqrt{3}-f(0)^NP(0,\dots,0)q_{(N)}&\nonumber\\ &\leq
(Q^{(N)},E_{expt}^{HV})&\nonumber\\
&\leq{2^{N-1}}\sqrt{3}-f(0)^NP(0,\dots,0)q_{(N)},
\label{row11}
\end{eqnarray}
where
\begin{equation}
q_{(N)}=\sum_{i_{1},\dots,i_{N}}Q_{i_{1},\dots,i_{N}}^{(N)}.
\label{row12}
\end{equation}
Therefore, since if $x$ is a possible value for $(Q^{(N)},E_{expt}^{HV})$ 
then so
is $-x$, one has
\begin{eqnarray}
&|({Q^{(N)}},E_{expt}^{HV})|\nonumber\\
&\leq{2^{N-1}}\sqrt{3}-P(0,\dots,0)|q_{(N)}|.
\label{a}
\end{eqnarray}
Thus, we have obtained Bell inequalities of a form which is more suitable
for the analysis of the experimental data.

The prediction (\ref{24}) leads to the following correlation function
\begin{equation}
E_{expt}^{QM}=\eta^{N}V(N)E^{QM},
\label{row13}
\end{equation}
which, when put into (\ref{a}) in the place of $E_{expt}^{HV}$, gives the 
following relation between the critical visibility, $V_{cr}(N)$,
and the critical collection efficiency, $\eta_N^{cr}$, for the $N$-particle
experiment:
\begin{equation}
{\eta_N^{cr}}^N\frac{3^N}{2}
V_{cr}(N)={2^{N-1}}\sqrt{3}-|q_{(N)}|(1-\eta_N^{cr})^N.
\label{c}
\end{equation}

The value of the expression $q_{(N)}$ can be found in the following way:
\begin{eqnarray}
&q_{(N)}=\sum_{i_{1},\dots,i_{N}}
\cos\left(\sum_{k=1}^{N}\phi_{i_{k}}\right)&\nonumber\\
&=Re\left(\sum_{i_{1},\dots,i_{N}}\prod_{k=1}^{N}\exp(i\phi_{i_{k}}^{k})
\right)&\nonumber\\
&=Re\left(\prod_{k=1}^{N}\sum_{i_{k}}\exp(i\phi_{i_{k}}^{k}\right)&
\nonumber\\
&=Re\left[2^{N}i\exp\left(i(N-1)
\frac{\pi}{3}\right)\right]=-2^{N}\sin\left((N-1)\frac{\pi}{3}\right).&
\end{eqnarray}

The critical value of the visibility of the multiparticle fringes decreases
now faster than in the earlier approaches \cite{MERMIN}. For perfect
collection efficiency, ($\eta=1$), it has the lowest value, which is
\begin{equation}
V_{cr}(N)={\sqrt{3}}(\frac{2}{3})^{N},
\end{equation}
and, if $N\geq4$, it is lower than $(\frac{1}{\sqrt{2}})^{N-1}$.  The
specific values for several particles are $V_{cr}(2)=77.8\%$,
$V_{cr}(3)=51.3\%$, $V_{cr}(4)=34.2\%$, $V_{cr}(5)=22.8\%$ and $V_{cr}(10)=3\%$,
whereas the standard methods lead to $V(2)^{old}=70.7\%$,
$V(3)^{old}=50.0\%$, $V(4)^{old}=35.4\%$, $V(5)^{old}= 25.0\%$ and
$V(10)^{old}=4.4\%$.  This suggests that for the original GHZ problem (four
particles) one should rather aim at making experiments which allow for three
settings at each local observation station.  Surprisingly, the measurements
should not be performed for the values for which we have perfect GHZ-EPR
correlations (i.e the values for which the correlation function equals to
$\pm1$)\cite{FOOT3}.

The critical efficiency of the particle collection also decreases with growing
$N$, and for perfect visibilities it reads $\eta^{(2)}=87.0\%$,
$\eta^{(3)}=79.8\%$, $\eta^{(4)}=76.5\%$, $\eta^{(5)}=74.4\%$. The gain over
the inequalities \cite{MERMIN} is in this respect very small, and begins
again at $N=4$.  However, for very big $N$ the critical efficiency is close
to $\frac{2}{3}$ (compared with $\frac{1}{\sqrt{2}}$ for \cite{MERMIN}). 

M.\ \.Zukowski acknowledges support of the University of Gda\'nsk research
grant no. BW-5400-5-0306-7, and of the 1996/97 Austrian-Polish
Scientific-Technological Collaboration Program PRO22, and a Visiting
Professorship of the University of Innsbruck.


\begin{references}
\bibitem{FOOT1}
We present our results only for the case of deterministic local 
hidden variable theories (for which $P_l(r_l|\lambda,\phi_l)=1$, or $0$),
but the generalization to stochastic theories is obvious.
\bibitem{FOOT3}
Seemingly, in the proposed inequalities 
altogether $3^4$ global settings are involved, but as 27 
elements the matrix (\ref{kukurydza8}), for $N=4$, vanish, thus 
effectively, if one  does not intend to 
make a delayed choice experiment, 
one may perform measurements at only 54 global settings.
\end{references}
\end{document}